\documentclass[prb,showpacs,citeautoscript, twocolumn]{revtex4}
\usepackage{graphicx}
\usepackage{amsmath}
\usepackage{amsbsy}
\usepackage{array} 

\begin{document}
\title{Electric Dipole Induced Spin Resonance in Disordered Semiconductors}

\author{Mathias Duckheim}
\email{mathias.duckheim@unibas.ch}
\author{Daniel Loss}
\email{daniel.loss@unibas.ch}
\affiliation{Department of Physics and Astronomy, University of Basel, CH-4056
Basel, Switzerland}

\begin{abstract}
  One of the hallmarks of spintronics is the control of magnetic moments by
  electric fields enabled by strong spin-orbit interaction (SOI) in
  semiconductors.  A powerful way of manipulating spins in such structures is
  electric dipole induced spin resonance (EDSR), where the radio-frequency
  fields driving the spins are electric, and not magnetic like in standard
  paramagnetic resonance.  Here, we present a theoretical study of EDSR for a
  two-dimensional electron gas in the presence of disorder where random
  impurities not only determine the electric resistance but also the spin
  dynamics via SOI.  Considering a specific geometry with the electric and
  magnetic fields parallel and in-plane, we show that the magnetization
  develops an out-of-plane component at resonance which survives the presence
  of disorder. We also discuss the spin Hall current generated by EDSR. These
  results are derived in a diagrammatic approach with the dominant effects
  coming from the spin vertex correction, and the optimal parameter regime for
  observation is identified.
\end{abstract}




\maketitle

The field of spintronics\cite{spintronics,Zutic} focuses on the interplay
between spin and charge degrees of freedom of the electron. The relativistic
effects responsible for the coupling between spin and orbital motion can be
strongly enhanced in solids due to band structure effects, with III-V
semiconductors showing a particularly strong spin-orbit interaction (SOI)
resulting in zero-field spin splittings. For instance, bulk inversion
asymmetry gives rise to Dresselhaus SOI\cite{dresselhaus:1955a}, while
structural inversion asymmetry occurring in heterostructures gives rise to
Rashba SOI\cite{rashba:1960a}.  The strength of such SOIs can be varied over a
wide range which offers the advantage to control magnetic moments with
electric fields.  A well-known and particularly powerful way of manipulating
spins in such structures is electric dipole induced spin resonance
(EDSR)\cite{bell_edt, rashba_combined,furdyna, merkt,awschalom3,rashba_inplane,schulte}, 
where the radio frequency (rf) fields coherently driving the
spins are electric, and not magnetic like in standard paramagnetic
resonance\cite{edelstein2,erlingsson}.  The advent of materials with tailored
SOI\cite{spintronics} has sparked intense interest in a variety of spin orbit
effects and its applications such as spin
currents\cite{dyakonov,murakami,sinova,awschalom,awschalom-she-2deg}, 
gate-controlled SOI effects\cite{salis:2001a,miller,rashba_orbit,kato},
spin relaxation in
quantum dots\cite{Elzerman,Kroutvar,Golovach},
zitterbewegung\cite{Schliemann}, spin-based quantum information
processing\cite{spintronics,loss:1998a}, and, in particular,
EDSR\cite{awschalom3,schulte,rashba_inplane}, which is the focus of this work.

Experimental indication for EDSR was recently reported for semiconductor
epilayers using an in-plane electric rf field\cite{awschalom3} (with the
magnetic field applied out of plane). In this geometry, spin coupling to the
electric field is much stronger than when the electric field is applied along
the growth direction\cite{rashba_inplane}.  Further, a recent experiment in
two-dimensional systems with cavities showed clearer signals for the
configuration with electric and magnetic fields perpendicular to each other, while in the parallel one the
observed resonance feature presented puzzles\cite{schulte}.

Systems of particular importance are heterostructures forming a
two-dimensional electron gas (2DEG), such as GaAs semiconductors.
Here, the Rashba SOI is linear in momentum and provides an effective internal
magnetic field about which the spin precesses.  Realistic 2DEGs, moreover,
contain disorder leading to momentum scattering which is responsible not only
for the finite electric resistance but also for spin relaxation due to
randomization of the internal field\cite{dyakonov1,dyakonov2}. 
Unlike for the conventional paramagnetic setting\cite{edelstein2} no
microscopic study of EDSR in such systems is available, but would be highly
desirable, also in view of the recent experimental activities. However, the
interplay between SOI and disorder in orbital space can be quite subtle, as was
fully appreciated only recently in a number of  experimental and theoretical 
studies on spin currents\cite{awschalom,awschalom-she-2deg,inoue,
mishenko, dimitrova, oleg}.  There, it turned out that the intrinsic spin
Hall effect\cite{sinova}  in GaAs 2DEGs due to
Rashba SOI does not survive the presence of disorder, which, technically
speaking, results from an unusual cancellation of vertex corrections\cite{inoue,
mishenko, dimitrova, oleg}.
Consequently, spin currents in these systems are
dominated by other (extrinsic) effects\cite{dyakonov,awschalom-she-2deg,engel}.  
Similar concerns, for instance, apply to conclusions reached for EDSR in clean
systems\cite{rashba_inplane}.  Thus, the role of disorder needs to be examined
carefully, and, in doing so, we show here that EDSR survives impurity
scattering but acquires a line shape that is determined by disorder and SOI.

\begin{figure}[t]
  \centering
\includegraphics[width = 8.5cm]{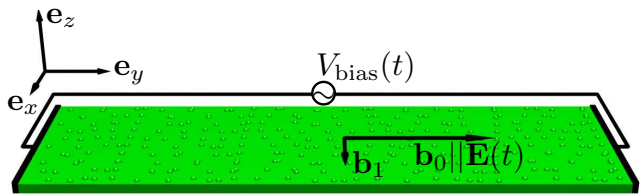}  
  \caption{{\bf Setup for electric dipole induced spin resonance (EDSR).} The
    disordered two-dimensional electron gas (2DEG) is placed in the xy-plane
    (green area). An a.c. bias $V_{\mathrm{bias}}(t)$ with frequency $\omega$
    generates an electric a.c. field $\mathbf E(t)$ along ${\bf e}_{y}$ which is
    in-plane and parallel to the externally applied magnetic field $\mathbf
    b_0 = g \mu_B \mathbf B_0 /2$ (assumed to be static). As a consequence of
    the orbital motion of the electron induced by $\mathbf E(t)$, the Rashba
    spin orbit interaction generates an 'internal' magnetic a.c. field $\mathbf
    b_1$ which is in-plane and perpendicular to the momentum of the electron
    and thus to $\mathbf b_0$ (ignoring disorder).  This configuration of
    fields is now analogous to conventional paramagnetic resonance and allows
    manipulation of the electron spin.
  }
  \label{fig:geometry}
\end{figure}

We consider a geometry as shown in Fig. \ref{fig:geometry}, where the electric
rf field and the static external magnetic field are parallel and both in the
plane of the 2DEG.  At resonance, the spin polarization (magnetization) acquires a non-zero
out-of-plane component.  This we show first for a clean 2DEG by deriving an
effective Bloch equation. Turning then to 2DEGs with disorder we treat the
electric rf field in linear response and obtain for the magnetization a
Lorentzian resonance whose width is given by a generalized D'yakonov-Perel
spin relaxation rate.  In addition, we find a shift of the resonance due to
disorder and SOI which gives rise to an effective g-factor that depends on the
magnetic field.  Using a standard diagrammatic approach to treat SOI and
disorder systematically, we find that it is the spin vertex correction (coming
from disorder) that leads to the resonance, in stark contrast to zero
frequency where the spin vertex vanishes\cite{edelstein}.  Assuming
realistic system parameters we identify the most promising regime for the
experimental observation of EDSR.  Finally, we discuss the spin Hall current and show
its relation to EDSR.

\section{The model}
The 2DEG consists of non-interacting electrons of mass m and
charge e which are subject to a random impurity potential $V$. In addition, we
allow for a general SOI $\mathbf \Omega(\mathbf p) \cdot  \boldsymbol{\sigma}$ linear in momentum
$\mathbf p$
and a static external magnetic field $\mathbf B_0$ applied
in-plane, as well as a time-dependent electric field $\mathbf E(t)$ applied as a
bias along $\mathbf B_0$, see Fig. \ref{fig:geometry}. The Hamiltonian for
this system reads
\begin{equation}
  \label{eq:h_ext}
  H = \frac{1}{2m} \left( \mathbf p - \frac{e}{c} \mathbf A \right)^2 +
  \mathbf \Omega ( \mathbf p) \cdot  \boldsymbol{\sigma}
  + (\mathbf b_0 +  \mathbf b_1(t) )\cdot \boldsymbol{\sigma} + V,  
\end{equation}
where $\mathbf b_1(t) =   - \frac{e}{c} \mathbf \Omega \left( \mathbf A(t)
\right)$ 
is the rf part of the internal 'magnetic' field induced by the electric 
field via SOI, and $\mathbf A(t) = -c \int^t dt' \mathbf E(t')$
is the associated vector potential, c being the speed of light. The Zeeman
term contains $\boldsymbol{b}_0 = g \mu_B \mathbf B_0 /2$ and the Pauli
matrices $\sigma^i$, $i = 1,2,3$, while $ \Omega(\mathbf p) $ is the
'zero-field spin splitting' due to the internal SOI field.

\section{Effective Bloch equation for the clean system} 
We show now that for the described setup 
the dynamics of the electron spin exhibits resonant behaviour (EDSR) generated by the electric rf field $\mathbf E(t)$.
 Starting with the simple case 
of no disorder ($V=0$), we 
derive a Bloch equation for the spin dynamics (for
weak SOI), from which the EDSR property immediately
follows. 
We begin by noting that the density matrix $\rho(t)$ is diagonal in momentum space,
and its elements can be expanded in the spin basis as
$\rho(\mathbf p,t) = \sum_{\nu=0}^{3}s^\nu(\mathbf p,t) \sigma^\nu$, where
$\sigma^0=1$. The expectation value of the spin  is then
$\langle \sigma^i (t) \rangle = \int d^2p \; s^i(\mathbf p, t) /(2\pi )^2
\equiv S^i(t)$, where $\int s^0 d^2p /(2\pi )^2 = 1$ due to normalization. 
[Henceforth, we refer to $S^i(t)$ as polarization, becoming the magnetization when multiplied
with the Bohr magneton $\mu_{B}$.]
In
momentum space the symmetry is broken by the small SOI term such that the
coefficients  decompose into an isotropic and a small anisotropic part,
$ s^i(\mathbf p,t) = \overline{ s}^i(p, t) + \Delta s^i(\mathbf
p,t)$. Averaging the
von Neumann equation for $\rho({\mathbf p},t)$ over directions of $\mathbf p$ we obtain
\begin{equation}
  \label{eq:spinbar}
  \dot{ \bar{\mathbf{s}}}(p,t) = \frac{2}{\hbar} (\mathbf b_0 + \mathbf b_1(t))  \times
  \mathbf{\bar s}(p,t)    ,
\end{equation}
where we have dropped the angular average of {$\mathbf \Omega(\mathbf p )
  \times (\overline{ \mathbf{ s}}(p,t) + \Delta \mathbf{ s}(\mathbf p,t) )$}
since it is higher order in the SOI.  Eq. ({\ref {eq:spinbar})
is now recognized as a  Bloch equation describing spin resonance.
Indeed, specializing henceforth to Rashba SOI $\Omega(\mathbf p) = \alpha
\mathbf p \times \mathbf e_z$, where $\mathbf e_z$ is a unit vector along the
confinement axis, and taking $\mathbf E$ in the plane along $\mathbf b_0 ||
\mathbf e_y$ results in the standard resonance setup\cite{Cohen-Tan} with an
oscillating internal field $\mathbf b_1(t) \propto \mathbf E \times \mathbf
e_z \perp \mathbf B_0$. Tuning $\mathbf E(t)$ to resonance, i.e.  to frequency
$\omega=\omega_{L}$, with $\omega_{L}=g\mu_{B}B_{0}$ being the Larmor
frequency, the spin starts to precess around the x-axis (in the yz-plane) with
a Rabi frequency $\omega_R = b_1/ \hbar =e E_{0} \alpha/ \hbar \omega_{L}$
given by the amplitude of the electric field $E_{0}$.

\section{Polarization of the Disordered 2DEG in Linear Response}
Having established the existence of EDSR for the clean system we turn now to
the realistic case of a disordered 2DEG.  For this we assume a dilute random
distribution of short-ranged scatterers with the disorder average
$\overline{V(\mathbf x) V(\mathbf x')} = (m \tau)^{-1} \delta(\mathbf x - \mathbf
x')$ taken to be $\delta$-correlated and proportional to the mean free time
$\tau$ between elastic scattering events. The interplay between SOI and
disorder can then be characterized by the dimensionless parameter $x=2 p_F
\alpha \tau/\hbar$ measuring the precession angle around the internal field $2
p_F \alpha /g\mu_{B}$ between scatterings.

For $A = 0$ and $V=0$ the eigenenergies of H become $E_s(\mathbf p) = p^2 / 2m
+ s\, b_{\mathrm{eff}}(\mathbf p)$, with $s= \pm 1$ and the effective magnetic
field $\mathbf b_{\mathrm{eff}} (\mathbf p ) = \alpha \mathbf p \times \mathbf
e_z + \mathbf b_0$. In the corresponding retarded (advanced) Green functions
\begin{equation}
\label{eq:green_function}
G^{R/A}(\mathbf p,E)=\frac{1}{2} \sum_{s=\pm1} \frac{1+s \frac{\mathbf
    b_{\mathrm{eff}} (\mathbf
    p)}{b_{\mathrm{eff}}(\mathbf p)} \cdot \boldsymbol{\sigma}}{E- E_s(\mathbf
    p) \pm i/ 2 \tau }   ,
\end{equation}
the disorder manifests itself as a finite self-energy term $i/2 \tau$
generated by the disorder average. In the self-consistent Born approximation
$\tau$ is independent of momentum due to the short range nature of $V$. We
have also checked that a renormalization of the Zeeman splitting, i.e. matrix
valued corrections, are of order $b_0/E_F$ and can be neglected as the Fermi
energy $E_F$ is taken to be the largest energy scale. Thus, the averaged Green
function is given by (\ref{eq:green_function}) with the standard isotropic and
spin independent term $i/2\tau$\cite{rammer}.

We turn now to the explicit calculation of the spin polarization (magnetization/$\mu_{B}$) $S^i(\omega)$
(per unit area) at frequency $\omega$, induced by the Fourier transform of the
electric field $\mathbf E(\omega)$. 
Working in the linear response regime we
start from the Kubo formula for $S^i(\omega)$ averaged over disorder and
evaluate it using standard diagrammatic techniques.  In Ref.  \cite{edelstein}
such a calculation was performed for the static ($\omega =0$) and zero field
($\mathbf B_{0}=0$) case, where, as a simplifying feature, the spin vertex
correction turned out to vanish.  For finite frequencies, however, this is no
longer the case, and, as we shall sketch now, it is this vertex correction
which leads to a finite out-of-plane polarization $S^3(\omega)$ at resonance
$\omega=\omega_{L}$.  To be specific, for $\hbar \omega \ll E_F$
the polarization becomes
\begin{equation}
  \label{eq:kubo}
  S^{i}(\omega) = \frac{-e E_j(\omega)}{2 \pi}  \mathrm{Tr} \left\{ G^A(q, E_F) \; \Sigma^{i}(\omega)  \: G^R(q, E_F +
  \omega) \; v_j \right\} ,
\end{equation}
where summation over repeated indices is implied. Here, the velocity operator
$v_j = (i/\hbar) [H, x_j] = p_j/m + \alpha ( \mathbf e_z \times
\boldsymbol{\sigma} )_j $ contains the usual spin-dependent term, and
$\mathrm{Tr} \rightarrow \int d^2 q /(2 \pi )^2 \,\mathrm{tr}$ denotes
momentum integration and tracing over spin states. In Eq.(\ref{eq:kubo}) we
introduced the spin vertex correction $\Sigma(\omega)$ determined by the
diagrammatic equation given in Fig. \ref{fig:renorm_vertex},
\begin{figure}[h]
  \centering
\includegraphics[width = 8.0cm]{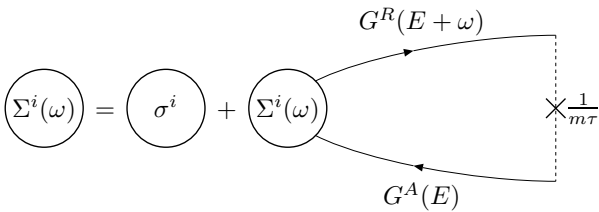}  
  \caption{{\bf Diagrammatic equation for the spin vertex correction $\Sigma(\omega)$.} 
    Iterating the equation for $\Sigma^{i}(\omega)$ generates the ladder
    diagrams accounting for the dominant contribution of the impurity
    scattering (denoted by a cross) to the Kubo formula Eq. (\ref{eq:kubo}).
    The impurities causing momentum scattering of the electron are assumed to
    be short-ranged, isotropic, spin-independent, and randomly distributed in
    the 2DEG.  Here, $\sigma^{i}$ denotes the Pauli matrix for the ith
    component of the electron spin, $G^{R/A}(E)$ is the retarded (advanced)
    Green function averaged over disorder (see Eq. \ref{eq:green_function}),
    and $ \tau $ is the mean-free-time between scattering events.}
     \label{fig:renorm_vertex}
\end{figure}

where the cross denotes the insertion of a factor $1/m \tau$.  The class of
diagrams generated by ${\Sigma}$ corresponds to the ladder approximation with
an accuracy of order $1/p_F l$, $l = p_F \tau /m$ being the mean free path
length. Thus, weak localization corrections\cite{edelstein1,oleg}, being
higher order in $1/p_F l$, are not considered here.

To further evaluate Eq.(\ref{eq:kubo}) we calculate $\Sigma$
in the (decomposed) form $\Sigma^i(\omega) \equiv \sum_{\nu = 0}^3 \Sigma^{i
  \nu}(\omega) \sigma^\nu$ for the limiting case of a strong magnetic field,
$b_0\gg \alpha p_F $, and for $\mathbf B_0=0$, respectively.
In both cases(cf. the methods section), the polarization is given by
\begin{equation}
\label{eq:mag1} 
  S^i(\omega) = e E_j(\omega)   \alpha \nu  2 \tau  \epsilon_{\mu j 3} \left[\delta^{i \mu} - \Sigma^{i \mu}(\omega) \left(1 -
  \frac{1}{\lambda}\right)\right]   ,
\end{equation}
where $\epsilon_{ijk}$ is the Levi-Civita tensor, $\lambda(\omega) = 1 - i
\omega \tau$, and $\nu=m/2 \pi \hbar^2$ denotes the two-dimensional density of
states. The form of $\Sigma(\omega)$, however, is to
be modified accordingly due to the presence (or absence) of the magnetic field
$\mathbf B_0$.

First, we consider the simpler case without magnetic field, i.e. $\mathbf
B_0=0$. Here, the off-diagonal elements $\Sigma^{\mu \nu}(\omega)$ vanish.
Thus, since the spin vertex is diagonal, we see that according to Eq.
(\ref{eq:mag1}) $\mathbf S(\omega)$ lies in-plane and perpendicular to
$\mathbf E(\omega)$, i.e. $S^3(\omega)$ vanishes identically for all $\omega$
in this case. In the considered geometry, with the electric field along the
y-axis, we are thus left with a finite in-plane polarization $S^1(\omega)$ in the
x-direction only. 
Further, for $\omega \to 0$ the factor $1-1/\lambda$
vanishes and thus the spin vertex $\Sigma^{i \mu}(0)$ drops out from Eq.
(\ref{eq:mag1}). This way, the result of \cite{edelstein} is recovered.

\begin{figure}[t]
  \centering
\includegraphics[width = 8.5cm]{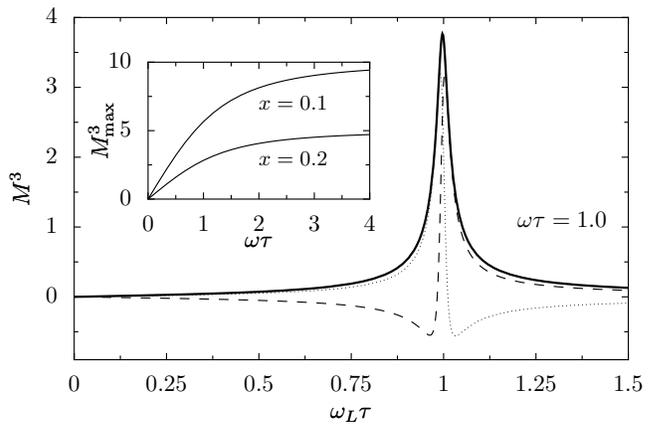}  
  \caption{{\bf EDSR and  linewidth in a disordered 2DEG.}
    The reduced out-of-plane polarization $M^3 = x (1-1/\lambda) \Sigma^{31}$ (cf.
    Eqs.(\ref{eq:mag1}), (\ref{eq:mag3trans}))  is plotted as a function of the
    Larmor frequency $\omega_{L}=g\mu_{B}B_{0}$, with resonance peak at
    $\omega_L = \omega$ of width $\Gamma$ given in Eq. (\ref{eq:damping2}).
    Modulus (full line), real (dashed) and imaginary part (dotted) are shown for
    $\omega \tau =1$ and $x=0.15$.  Inset: Peak height of $M^3$ at resonance
    as a function of $\omega \tau$ for $x=0.2$ and $x=0.1$. }
  \label{fig:resonance}
\end{figure}

Next we consider the opposite case of a magnetic field $\mathbf B_0$ that is
large compared to the internal SOI field. We characterize this regime by the
small expansion parameter $a = \alpha p_F/ 2 b_0 = x/2 \omega_L \tau\ll 1$,
being the ratio of precession angles around the internal and external magnetic
fields, respectively, between scatterings. The field $\mathbf B_0$ leads to an
equilibrium polarization along the y- axis due to Pauli paramagnetism given by
$S_{\mathrm{eq}} = \nu g \mu_B B_0/2$. As we shall see, 
the EDSR response generated by the electric field $\mathbf E(t)$ 
will be proportional to $S_{\mathrm{eq}}$, and thus a finite equilibrium magnetization
is crucial. [We note  that  the Kubo formula, Eq.
(\ref{eq:kubo}), describes  the {\it deviation} from equilibrium only in
response to the electric (and not the magnetic) field.]

With both, the internal and external magnetic fields present, the dispersion
relation $E_s(\mathbf p)$ is no longer isotropic as $ b_{\mathrm{eff}}$
depends on the direction of $\mathbf p$. This complicates the momentum
integrations considerably, and as a further complication, 
$\Sigma$ becomes off-diagonal through the additional spin terms in the Green
functions. Fortunately, for $a\ll 1$ analytical progress is still possible,
and $\Sigma(\omega)$ can be evaluated explicitly as we outline in the methods
section.

The polarization $S^{i}(\omega)$ is again given by Eq. (\ref{eq:mag1}) but now
with the spin vertex found for $B_0 \neq 0$ (see Eq. (\ref{eq:sigma1}) below),
which has the finite off-diagonal elements $\Sigma^{13} = - \Sigma^{31}$.
Accordingly, the out-of-plane component $S^{3}$ is proportional to the
reduced magnetization $M^{3} \equiv x\Sigma^{31} (1 - 1/\lambda)$ which is
plotted in Fig.  \ref{fig:resonance} as a function of the Larmor frequency
$\omega_{L}$, exhibiting a pronounced peak at the resonance
$\omega_{L}=\omega$ (up to a small shift, see below). This clearly shows that
the EDSR resonance found in the clean case survives weak disorder, and,
technically speaking, stems from the spin vertex correction.

\section{EDSR and linewidth}
Let us now analyze the line shape of the peaks close to resonance $\omega_{L}
= \pm \omega $ in more detail, i.e. for $ |\omega_{L} + \omega| \ll
|\omega_{L} - \omega|$ (left) and $ |\omega_{L} - \omega| \ll |\omega_{L} +
\omega|$ (right peak), respectively. In this case and to order $x^2$ Eq.
(\ref{eq:mag1}) can be rewritten as a sum of two Lorentzians, with the
transverse components ($i = 1,3$) becoming
\begin{align}
  \label{eq:mag3trans}
  &S^i( \omega) = e E_y(\omega) \alpha \nu  \frac{ \omega_i
  \tau}{1 - i \omega \tau}    \notag \\ &\times \left( \frac{1}{\omega_L -
  \omega +\delta \omega  
-i\Gamma} + \frac{1}{\omega_L + \omega -\delta \omega
  +i\Gamma} \right)     ,
\end{align}
where we denote $\omega_1 \equiv \omega_L$ for the in-plane ($i=1$) and
$\omega_3 \equiv -i \omega $ for the out-of-plane component ($i=3$),
respectively. Analogously to conventional paramagnetic resonance, the prefactor
of $S^i$ can be rewritten in terms of the Rabi frequency $\omega_R = e E_{0}
\alpha/ \hbar \omega_L$, i.e., the amplitude of the 'internal' rf field $\bf b_{1}(t)$, and the
equilibrium polarization $S_{\mathrm{eq}} = \nu g \mu_B B_0/2$ due to Pauli paramagnetism.

Eq. (\ref{eq:mag3trans}) is the main result of this work. Here, $\Gamma$ is
the linewidth of the resonance peak and is explicitly given by the imaginary
part of the damping function Eq. (\ref{eq:damping})
evaluated at the (bare) resonance,
\begin{equation}
 \label{eq:damping2}
\Gamma = \frac{x^2}{ 2\omega_{L} \tau^{2}} \, \mathrm{Im} \; \gamma (\omega=\omega_{L})=
\frac{x^2}{ 2\tau} \left( 1 +  \frac{1}{2[1 + (\omega_{L}
 \tau)^2]}     \right).
\end{equation}
Here, the prefactor $x^2/2 \tau$ is the familiar D'yakonov-Perel spin
relaxation rate\cite{dyakonov1, dyakonov2} coming from internal random fields
induced by disorder.  The second factor accounts for the magnetic field. We
note that $\Gamma$ saturates for $\omega_{L}\tau \gg 1$ at the finite value
$x^2/ 2\tau$, and, thus, dynamical narrowing\cite{Abragam, rashba_combined} due to fast spin
precession is incomplete.  This reflects the SU(2) nature of the spin with two
precession axes on the Bloch sphere, one given by the external and one by the
internal field. At resonance and for $\omega_{L}\tau \gg 1$ the dynamical
narrowing effect due to the external field is complete, but not the one due to
the internal field, since in the linear response regime considered here the Rabi
frequency is restricted to the regime $\omega_{R}\tau\ll1$ (see appendix).  Thus, in
order to describe complete dynamical narrowing in the presence of disorder one
needs to go beyond linear response, which, however, is outside the scope of
the present work. For sufficiently strong E-fields in the limit
$\omega_{R}\tau\gg1$ it is plausible to expect that disorder becomes
irrelevant for the spin dynamics. In this case, the Bloch equation
(\ref{eq:spinbar}) applies and, like in standard paramagnetic resonance, the
linewidth will then be given by the internal field $b_{1}$ (see also Sec.\ref{sec:validity}).

Next, the bare location $\omega_{L} $ of the resonance gets renormalized by
the disorder leading to a small shift $\delta \omega\ll \omega_{L}$,
determined by the real part of $\gamma$ in Eq. (\ref{eq:sigma1}),
\begin{equation}
  \label{eq:damping1}
\delta \omega= \frac{x^2}{ 2\omega_{L} \tau^{2}}\mathrm{Re} \;  \gamma (\omega=\omega_{L})= 
  \frac{\omega_{L} x^{2}/4}{1 + (\omega_{L}
 \tau)^2}.
\end{equation}
We note that for $\omega_{L}\tau \sim 1$ we get $\Gamma \sim 5\delta\omega$.
Also, the shift $\delta\omega$ depends non-monotonically on the magnetic field
($\omega_{L}$), giving rise to an effective g-factor, $g_{eff} =g
(1+x^{2}/4({1 + (\omega_{L}\tau)^2}))$, which is B-field dependent.
Observation of this effect in real measurements would provide useful
additional evidence for EDSR. Finally, the longitudinal component (i.e. along
$\bf B_{0}$) of the polarization vanishes identically, i.e., $S^2( \omega)
=0$.  This is not quite unexpected in the linear response approximation, since it
is known from conventional paramagnetic resonance\cite{Cohen-Tan} that changes
in the occupation probability are nonlinear in the driving field.

Let us now give some numbers to illustrate the EDSR effects discussed here
(for details see Sec. \ref{sec:estimates}). To quantify the amount of spins resonantly excited, we
introduce the relative polarization $P = (N_\uparrow - N_\downarrow)/(N_\uparrow +
N_\downarrow) $ given by the ratio of $S^3(\omega)$ and the electron sheet
density.  Here $N_\uparrow$ and $N_\downarrow$ denote the average number of
spins pointing up and down along the $z-$axis, respectively. Choosing now
parameter values typical for a GaAs 2DEG sample of a few hundred micrometers
in lateral size, a magnetic field of one Tesla, and an a.c. bias of 0.1 Volt and
8 GHz we get $P= 10^{-4}$ at resonance.  This corresponds to $N_\uparrow -
N_\downarrow = 200$ excess spins in the laser spot of a typical optical
measurement scheme\cite{awschalom-she-2deg}, and although being small, this is within
reach of detection. Furthermore, the associated  
Rabi frequency, relaxation rate, and resonance shift, respectively, become
$\omega_R/2\pi =12$ MHz, $\Gamma/2\pi = 50$ MHz, and $\delta \omega/2\pi =10$ MHz.
We note that the 
polarization $P$ increases at resonance with increasing frequency $\omega$
(cf. inset in Fig. \ref{fig:resonance} and Eqs. (\ref{eq:damping2}) and
(\ref{eq:mag3trans})). Thus, the higher the resonance frequency (and the
magnetic field) the larger the EDSR signal.

\section{Spin Hall current}
We finally turn to a brief discussion of the associated spin Hall current for
a homogeneous infinite 2DEG (for details see Sec. \ref{sec:spin-current}). Using the Heisenberg
equation of motion we can express the spin Hall current, defined as $I^3_x =
\langle \{ \sigma^3, v_x \} \rangle /2$\cite{sinova}, in terms of the transverse polarization
components as
\begin{align}
  \label{eq:spincurrenta}
  I^3_x(\omega) = \frac{\hbar}{2 \alpha m } \left[ i \omega S^1 (\omega) +
  \omega_{L}S^3(\omega) \right].
\end{align} 
The importance of this relation lies in the fact that it establishes a direct link between the spin Hall current
and the spin polarization, and that the latter quantity is experimentally accessible by known measurement
techniques.
 We note further that for $I^3_1
\equiv 0$  Eq. (\ref{eq:spincurrenta}) would describe  circular motion of $S^{1}$ and $S^{3}$
 in the x-z plane (at resonance), i.e.  'pure' precession of the spin around the y-axis. 
 A finite spin current, therefore,
characterizes the deviation from pure  precession   and gives rise to nutation
 of the spin.
From the ratio $\hbar I^3_x /2 E_{y}$ the associated a.c. spin Hall conductivity
follows, which, at resonance, is explicitly given by (see Eq. (\ref{eq:sh-conductivity})) $ \sigma^{3,
  res}_{x y}= (e/4 \pi) i \omega_L \tau/(1 + 2 \lambda)$.  Thus, for the high
field regime $\omega_{L}\tau \gg 1$ we see that the spin Hall conductivity
assumes a universal value $ \sigma^{3, res}_{x y} = |e|/ 8 \pi$. Quite
remarkably, this is the same value as obtained in the {\it absence}
of magnetic fields and at finite frequencies\cite{mishenko}. 

In conclusion, we have investigated the spin polarization of a 2DEG in the
presence of spin-orbit interaction and disorder. We have shown that carrier
spins in a specific field configuration with an electric rf field give rise to
EDSR with a linewidth coming from spin relaxation due to disorder and
spin-orbit interaction.  Our results emphasize the importance of tunable SOI
for coherent spin manipulation by electric means in semiconductors.\\

{\bf Acknowledgments}
\\We thank O. Chalaev, J. Lehmann, D. Bulaev, W. Coish, S. Erlingsson, D. Saraga, and
D. Klauser for discussions. This work was supported by the Swiss NSF, the NCCR
Nanoscience, EU RTN Spintronics,
DARPA, and ONR.\\

\appendix

\section{Methods}
Here we sketch some of the important steps of the calculation. A more detailed
account can be found in Sec. \ref{sec:spinvertex}.

{\bf Calculation of the spin vertex correction.}
For the evaluation of Eq. (\ref{eq:kubo}) we introduce the spin-spin
($X^{\mu \nu}$) and the spin-momentum ($Y^{\mu j}$) terms,
resulting from the spin and the normal part of $v_j$, respectively,
\begin{align}
  \label{eq:xy}
  X^{\mu \nu}(\omega) &= \frac{1}{2 m \tau}  \mathrm{Tr} \left\{ G^A(q, E_F) \; \sigma^{\mu}  \: G^R(q, E_F + \hbar
  \omega) \; \sigma^{\nu} \right\} \notag \\
  Y^{\mu j}(\omega) &= \frac{1}{2 m \tau} \mathrm{Tr}  \left\{ G^A(q, E_F) \; \sigma^{\mu}  \: G^R(q, E_F + \hbar
  \omega) \; \frac{p_j}{m} \right\}   
\end{align}
and expand the spin vertex in terms of Pauli matrices,
$\Sigma^i(\omega) = \sum_{\nu=0}^3 \Sigma^{i \nu}(\omega) \sigma^\nu$ with the coefficients
$\Sigma^{\mu \nu}$ being represented by a $4 \times 4$-matrix. 
Here, the Fourier transform of a function $f(t)$ is defined as
$f(\omega)=\int_{-\infty}^{+\infty}dt f(t) e^{i\omega t}/2\pi$.
With this
notation we then obtain from Eq. (\ref{eq:kubo}) for the polarization
\begin{equation}
  \label{eq:mag}
  S^i(\omega) = \frac{-e E_j(\omega)  m \tau}{ \pi \hbar^2}
  \Sigma^{i \mu}(\omega) \left( \alpha X^{\mu \nu}(\omega) \epsilon_{ \nu j 3}  + Y^{\mu j}(\omega)\right).
\end{equation}

We proceed using the identity $\delta^{\alpha \alpha'} \delta^{\beta \beta'} = 1/2
\sum_{\mu=0}^3 \sigma^\mu_{\alpha \beta} \sigma^\mu_{\beta' \alpha'}$ such
that the diagrammatic equation in Fig. \ref{fig:renorm_vertex} factorizes and can be
solved. This way, the matrix elements of the ``diffuson'' are obtained in the form
${\Sigma^{\mu \nu}}(\omega) = [\left[ \mathbf 1 - {X}(\omega)\right]^{-1}]^{\mu
\nu}$ and the polarization $S^i(\omega)$ is
expressed in terms of the $4 \times 4$-matrices ${Y}$, ${X}$, and ${\Sigma}$,
which were evaluated for the two cases $\mathbf
b_0=0$ and $b_0\gg \alpha p_F $, respectively.

{\bf The case $\mathbf B_0 = 0$.}
Performing the integrals in Eq. (\ref{eq:xy}) for $\mathbf B_0 = 0$ we find
$X^{00}(\omega) = 1/\lambda$ and $X^{33}(\omega) = \lambda/(\lambda^2 + x^2)$,
while the in-plane components are equal, $X^{11}= X^{22} = (X^{00} + X^{33})/2
$ (these results have already been obtained in \cite{oleg} in the evaluation
of the
spin Hall conductivity).  
The off-diagonal components $X^{\mu\nu}$
vanish so that the spin vertex follows simply as $\Sigma^{\mu \mu} = (1 -
X^{\mu \mu})^{-1}$.  The spin-momentum matrix is found as $Y^{\mu j}= -\alpha
\epsilon_{\mu j 3} / \lambda +O(\alpha^{3})$ (see also Sec. \ref{sec:spinvertex}).

{\bf The case $\mathbf B_0 \neq 0$.}
In the case of the strong magnetic field, i.e., $\mathbf B_0 \neq 0$ it turns
out that, in leading order in the SOI-strength $\alpha$ the spin-momentum term
keeps the same form as in the absence of the magnetic field, i.e. $Y^{\mu j}=
-\alpha \epsilon_{\mu j 3} / \lambda$. The polarization is thus again
described by Eq. (\ref{eq:mag1}), however, with the spin vertex modified by
the presence of the field $\mathbf B_0$ (cf.  Eq. (\ref{eq:sigma1})). In this
case, the calculation of $\Sigma(\omega)$ is more involved since, starting
from weak SOI-strength $a \ll 1$, the leading order power in $a$ changes
precisely at resonance. Thus, an expansion up to second order in $a$ of the
spin-spin term $X$ with the subsequent matrix inversion $\Sigma =
(1-X)^{-1}$ is required to consistently describe the shape of $S^i$ in terms
of the spin vertex at resonance. 

As a result (see Sec. \ref{sec:spinvertex}), we obtain
\begin{align}
  \label{eq:sigma1}
  &{\Sigma} (\omega)= \notag \\
&\left( 
    \begin{array}[c]{cccc}
       \frac{\lambda}{\lambda -1} & 0 & 0 & 0 \\ 
      0 &  \frac{\lambda^2 + (\omega_L \tau)^2 - \lambda + x^2 \gamma_{11}}{\tau ^2 (\omega_L^2
        - \omega^2)  + x^2  \gamma} & 0 &  \frac{ \omega_L \tau + x^2
      \gamma_{13}}{\tau ^2 (\omega_L^2
      - \omega^2)  + x^2  \gamma} \\
    0 & 0 &  \frac{\lambda}{\lambda -1} & 0 \\
    0 &  - \frac{ \omega_L \tau + x^2 \gamma_{13}}{\tau ^2 (\omega_L^2
      - \omega^2)  + x^2  \gamma}  & 0 &   \frac{\lambda^2 + (\omega_L \tau)^2 - \lambda + x^2 \gamma_{33}}{\tau ^2 (\omega_L^2
        - \omega^2)  + x^2  \gamma}
    \end{array}
\right)   ,
\end{align}
where the imaginary part of the complex  function
\begin{equation}
  \label{eq:damping}
  \gamma( \omega) = \frac{3 (\lambda -1) \lambda^3  -
  (\omega_L \tau)^2 \lambda (6 \lambda -1)-(\omega_L \tau)^4}{2 \lambda (\lambda^2 + (\omega_L
  \tau)^2)^{2}} 
\end{equation}
characterizes the linewidth induced by the impurity scattering via the SOI, as
described in the main text. The functions $x^2 \gamma_{11}(\omega)$ and $x^2
\gamma_{13}(\omega)$ (explicitly given in Eq. (\ref{eq:x2func})) are negligible for obtaining the
polarization $S^{i}(\omega)$. However, for obtaining the spin Hall current
they are needed since the the leading terms of $\Sigma^{11}$ and $\Sigma^{31}$
cancel each other in Eq. (\ref{eq:spincurrent}).

\section{Calculation of the  spin vertex correction} \label{sec:spinvertex}
Here, we give a more detailed account of the evaluation of the polarization
$S^i(\omega) $, Eq. (\ref{eq:mag}), in terms
of the spin-spin and the spin-momentum terms, $X$, $Y$, respectively, and
the  spin vertex correction $\Sigma$. The simultaneous presence of the
internal and external fields, $\alpha (\mathbf p \times \mathbf e_z)$ and
$\mathbf b_0$, resp., breaks the symmetry in orbital space
such that no closed analytical expression for the
integrals in Eq. (\ref{eq:xy}) and hence for $X^{\mu \nu}$  can be obtained in the general case.
However, the most important regime for EDSR is given by the regime where the
internal field is much smaller than the (perpendicular) static external
magnetic field, as in standard paramagnetic resonance (see also next section).
Thus, without any essential restriction we can concentrate on the regime with
the SOI being small compared to $\mathbf b_0$, i.e.  $a =\alpha p_F/2 b_0 = x
/ 2 \omega_L \tau \ll 1 $.  First, upon inspection of Eq. (\ref{eq:mag}) we note that
the contribution of $Y$ to the polarization is due to the momentum-part of the
velocity and thus must vanish in the absence of SOI. Thus, the leading order
term in Eq. (\ref{eq:mag}) coming from $Y$ is at least linear in $a$ (assuming
analyticity). More precisely, with a calculation similar to the one outlined
below for $X$, we obtain for the spin-momentum diagram
\begin{equation}
  \label{eq:spin-momentum}
Y^{\mu j}= -\epsilon_{\mu j 3} \frac{\alpha}{\lambda(\omega)}+ O(\alpha^{3}),
\end{equation}
i.e. the same result as before for $b_0 = 0$. Note that only odd powers in
$\alpha$ appear due to the symmetry constraints induced by the angular
integration occurring in $Y$.  Thus, the expression Eq. (\ref{eq:mag1}) is linear in the
SOI $\alpha$ in leading order.  In order to expand the polarization Eq. (\ref{eq:mag1})
in leading order in $\alpha \propto a$ it is therefore sufficient to calculate
the spin-spin diagram with setting $\alpha$ to zero and retaining only the
$\mathbf b_0$ dependence. This way, we obtain the spin vertex correction
$\Sigma(\omega)$ which is singular at resonance, i.e. when $\omega=\omega_L $
(Larmor frequency), reflecting the presence of Rabi oscillations. This shows,
however, that at resonance the next-to-leading order contributions of the
spin-spin diagram $X = X_{(0)} + a^2 X_{(2)}$ become relevant\footnote{As a
  general property of linear SOI the first order in $a$ vanishes due to the
  symmetry in the angular integration. Indeed, we note that the angular
  dependence (in the integrals in Eq. (\ref{eq:xy})) comes from the magnetic field (Eq.
  (\ref{eq:magfield})) where $\varphi$ always occurs in terms of a
  trigonometric function simultaneously with a factor $a$. Expanded in $a$ the
  linear terms thus vanish upon angular integration.} for the matrix inversion
and must be kept.  Indeed, they represent the dominant contribution in $
\Sigma(\omega) = (\mathbf 1 - X_{(0)}(\omega) - a^2 X_{(2)}(\omega) )^{-1}$ if
the determinant of the first term $\boldsymbol{1} - X_{(0)}$ vanishes.
Obviously, at resonance the dominant $a$-dependence of $\Sigma$ becomes
$1/a^2$.
Hence, we concentrate on the evaluation of
the spin-spin diagram up to  order of $a^2$, with $
X_{(2)}$ characterizing the behavior of the polarization around the
resonance (where our analysis is valid).

The spin-spin diagram is given by
\begin{align}
  \label{eq:spin-spin}
  X^{\mu \nu} & = \frac{1}{4 E_F \tau}  \sum_{s, s' = \pm 1} \int \frac{d^2
  Q}{(2 \pi)^2} T_{s, s'}^{\mu \nu} (\mathbf Q)
  \\ & \times \frac{1 }{(1 + w - Q^2 - s B(\mathbf Q) +
  i r/2)} \notag \\ & \times \frac{1}{ (1 - Q^2 - s' B(\mathbf Q) - i r/2)}\notag
\end{align}
expressed in terms of the dimensionless quantities\footnote{In this
  Section \ref{sec:spinvertex} the capital letters $\mathbf B, \mathbf B_0$ and
  $B, B_0$ denote dimensionless magnetic fields measured in units of the Fermi
  energy $E_{F}$. } $w = \hbar \omega/E_F$, $r=\hbar / E_F \tau$, $Q= q/p_F$,
$B_0 = b_0/E_F$ and the effective magnetic field
\begin{equation}
  \label{eq:magfield}
  \mathbf B(\mathbf Q) = \frac{\mathbf b_{\mathrm{eff}} (\mathbf Q \,p_F)}{E_F} = \mathbf B_0 + 2 a B_0
  (\mathbf Q \times \mathbf e_z) 
\end{equation}
with modulus
\begin{equation}
  \label{eq:magfield_mod}
  B(\mathbf Q) = B_0 \sqrt{ 1+ 4 a \, \hat{\mathbf{B}}_0 \cdot (\mathbf Q \times
  \mathbf e_z) + 4 a^2 Q^2 }.
\end{equation}
Here, $\hat{\mathbf{B}}_0 = \mathbf B_0 / B_0$ is the unit vector of the
external field taken
along the $y$-axis such that the
mixed product in Eq.(\ref{eq:magfield_mod}) becomes $\hat{\mathbf{B}}_0 \cdot (\mathbf Q \times
\mathbf e_z) =  - Q \cos \varphi$. 

{\bf Trace.}  In Eq.(\ref{eq:spin-spin}) we introduced the trace over spin
states 
\begin{align}
  \label{eq:spintrace}
T_{s,s'}^{\mu \nu}(\mathbf Q) &=  \mathrm{tr} \{ \sigma^\mu (1+s \,
\hat{\mathbf{B}}(\mathbf Q)  \cdot \boldsymbol{\sigma}) \sigma^\nu(1+s' \,
\hat{\mathbf{B}}(\mathbf Q) \cdot  \boldsymbol{\sigma})\} \notag \\
&= 4 \,  \delta^{\mu \nu } [\delta^{\mu  0}\delta_{s,s'}  + 
\delta^{\mu \neq 0}   \delta_{s,-s'}] \notag \\
&+ 4 \, [\delta^{\mu \neq 0} \delta^{\nu 0} \hat B_{\mu}(\mathbf Q) +
\delta^{\mu  0} \delta^{\nu \neq 0} \hat B_{\nu} (\mathbf Q) ]
 s \delta_{s,s'} \notag \\ & +   4 \, i \epsilon_{\mu \nu k} \hat B_k(\mathbf Q) 
\delta^{\mu \neq 0} \delta^{\nu \neq 0} s  \delta_{s,-s'} \notag \\
&+ 4 \, s \, s' \hat B_\mu (\mathbf Q) \hat B_\nu (\mathbf Q)
\delta^{\mu \neq 0} \delta^{\nu \neq 0} \,\, ,
\end{align}
where $\delta^{\mu \neq 0}=1-\delta^{\mu  0}$ etc., and where summation over repeated indices is
implied. There are 
%
%
terms containing none, one, or two normalized magnetic fields 
$\hat{\mathbf{B}} (\mathbf Q) = \mathbf{B} (\mathbf Q) / B (\mathbf Q)$, which
is relevant for the momentum integration. The trace $T_{s,s'}^{\mu \nu}(\mathbf
Q)$ and the direction of $\mathbf B_0$ determine the matrix structure of the spin-spin
diagram, i.e. which components $X^{\mu \nu}$ are nonzero. 

{\bf Momentum integration.}
\begin{figure}[t]
  \centering
\includegraphics[width = 8.5cm]{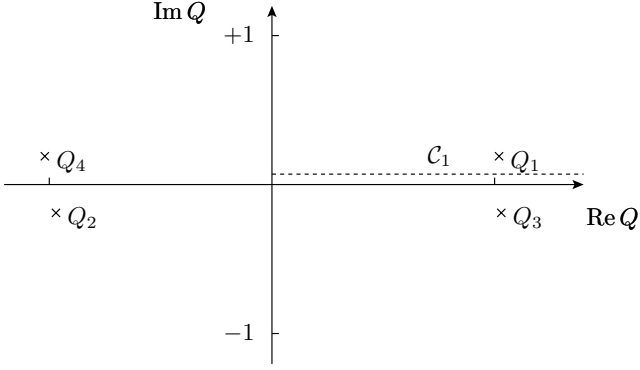}  
  \caption{View of the poles $Q_i$ of the retarded and advanced Green functions in the
    complex plane (cf. Eq. ({\ref{eq:poles}}) ). The contour $\mathcal{C}_1$ of
    the momentum integral (Eq. (\ref{eq:spin-spin1})) running from $0$ to $+
    \infty$ is shown.}
  \label{fig:poles}
\end{figure}
The components $X^{\mu \nu}$ are obtained by the momentum integration in
Eq. (\ref{eq:spin-spin}) where the poles $Q_i$ of the denominator yield the
main contribution. Assuming that $E_F$ represents the largest
energy scale such that $w$, $r$, and $B_0$ are small compared to one, the poles are
located essentially at $|Q| \approx 1$ (cf. the denominator in Eq.
(\ref{eq:spin-spin})) with corrections $O(w, B_0, r)$. Making use of $a \ll 1$
we expand the denominator of the integrand in Eq. (\ref{eq:spin-spin}) in $a$
up to second order. From this we obtain the following four poles
\begin{align}
  \label{eq:poles}
Q_1 &= +1 + k_1 + i \frac{r}{2} \notag \\
Q_2 &= -1 + k_2 - i \frac{r}{2} \notag \\
Q_3 &= +1 + k_3 - i \frac{r}{2} \notag \\
Q_4 &= -1 + k_4 + i \frac{r}{2} \, \, ,
\end{align}  
where
\begin{align}
k_{1, 2} &= \pm \frac{w}{2} \mp s \frac{B_0}{2} + B_0 a s \cos \varphi \mp a^2
s B_0 \sin^2\varphi\notag \\
k_{3, 4} &= \mp s' \frac{B_0}{2} +B_0 a s' \cos \varphi \mp a^2 s' B_0 \sin^2
\varphi \,\, .
\end{align}
These poles are approximately of order one with a small correction $k_i$ and a small imaginary
part $ r/2=\hbar/2E_{F}\tau$, thus showing that the above approximation for small $a$ is
self-consistent. Decomposing into linear factors the spin-spin diagram  can be
recast into the form
\begin{align}
\label{eq:spin-spin1}
X^{\mu \nu} = \frac{1}{8 \pi  E_F \tau} \sum_{s,s'= \pm 1} \langle
\int_0^{+\infty} \frac{dQ \,Q \,T^{\mu \nu}_{s,s'} (Q) }{(Q - Q_1) \dots (Q - Q_4) c_\varphi}
\rangle_{\varphi}  \,\, , 
\end{align}
where $\langle ...\rangle_{\varphi}$ denotes integration over the polar angle (normalized
by $2 \pi$), and $c_\varphi = 1 + a^2 B_0 (s + s') \sin ^2
\varphi$.

Note that the $Q$-dependence of $1/B(Q)$ in $T_{s,s'}^{\mu \nu}$ for $a \neq 0$ generates additional
singularities at $Q_{5/6} = \exp{(\pm i \varphi)} / 2 a$ when analytically continued
into the complex plane away from the real axis. The application of complex contour
integration is thus non-trivial. A direct calculation, however, carried out by
decomposing the denominator in Eq.(\ref{eq:spin-spin1}) into partial fractions
shows that these poles do not contribute to $X$ within the accuracy $O(w, B_0, r)$.

The subsequent angular integrations become simple when expanding in the
small parameter $a$. In this way, we obtain the components $X^{\mu \nu}$ and via
the matrix inversion $\Sigma = (\boldsymbol{1} - X)^{-1}$ the 
spin vertex correction in Eq. (\ref{eq:sigma1}) of the paper. 
In particular, the vertex components
\begin{align}
  \label{eq:sigma_comp}
\Sigma^{11} &= \frac{(\omega_L \tau)^2 + (\lambda-1) \lambda + x^2
  \gamma_{11}}{(\omega_L \tau)^2 - (\omega \tau)^2 + x^2 \gamma} \notag \\
\Sigma^{13} &= -\Sigma^{31} =\frac{\omega_L \tau +  x^2
  \gamma_{13}}{(\omega_L \tau)^2 - (\omega \tau)^2 + x^2 \gamma}
\end{align}
with the complex damping function
\begin{equation}
  \label{eq:dampinga}
  \gamma( \omega) = \frac{3 (\lambda -1) \lambda^3  -
  (\omega_L \tau)^2 \lambda (6 \lambda -1)-(\omega_L \tau)^4}{2 \lambda (\lambda^2 + (\omega_L
  \tau)^2)^{2}} 
\end{equation}
characterizing the linewidth and the functions
\begin{align}
  \label{eq:x2func}
\gamma_{11}(\omega, \omega_L) &= -\frac{\lambda[ (\omega_L \tau)^2 - \lambda^2]}{[(\omega_L \tau)^2 +
  \lambda^2]^2}  \notag \\
\gamma_{13}(\omega, \omega_L) &= -\frac{2 \omega_L \tau \lambda^2}{[(\omega_L
  \tau)^2 + \lambda^2]^2}
\end{align}
are relevant for the subsequent calculation of the spin polarization and spin Hall current.
The frequency dependence and resonance behaviour of the spin
polarization and current are discussed in the main text.

\section{Regime of validity}\label{sec:validity}
We now give a summary of the parameters controlling the regime of validity of the present
theory.

A first constraint ensures the validity of the linear response approach.  For
this, we give a heuristic argument based on the analogy to conventional
ESR\cite{Cohen-Tan} expressed by Eq. (\ref{eq:spinbar}) of the paper.
For this case, we consider the Bloch equations
\begin{align}
  \label{eq:bloch}
\dot{S}^1  &= \gamma \left[ \mathbf S \times (\mathbf B_0 + \mathbf{B}_1(t) ) \right]_1 -
\frac{S^1 }{T_2} \notag \\
\dot{S}^3  &= \gamma \left[ \mathbf S \times (\mathbf B_0 + \mathbf{B}_1(t) ) \right]_3 -
\frac{S^3}{T_2} \notag \\
\dot{S}^2  &= \gamma \left[ \mathbf S \times (\mathbf B_0 + \mathbf{B}_1(t) ) \right]_2 -
\frac{S^2  - S_{eq} }{T_1} 
\end{align}
describing magnetic moments subject to a constant magnetic field $\mathbf B_0
|| \mathbf e_2$ and a circularly polarized field $\mathbf{B}_1(t)$
perpendicular to it, which oscillates with frequency $\omega$. The familiar steady
state solutions in the rotating frame for the longitudinal component $S^2$ and
the transverse components $S^u$ and $S^v$, respectively are
\begin{align}
  \label{eq:bloch:sol}
S^2 &= S_{\mathrm{eq}} \frac{1 + \Delta \omega^2 T_2^2}{1 + \Delta \omega^2 T_2^2 + T_1
  T_2 \omega_1^2 } \notag \\
S^u &= S_{\mathrm{eq}} \frac{\Delta \omega \omega_1  T_2^2}{1 + \Delta \omega^2 T_2^2 + T_1
  T_2 \omega_1^2 } \notag \\
S^v &= S_{\mathrm{eq}} \frac{\omega_1 T_2}{1 + \Delta \omega^2 T_2^2 + T_1
  T_2 \omega_1^2 } ,
\end{align}
where $\omega_1 = \gamma B_1$, $\omega_L = \gamma B_0$, and $\Delta \omega =
\omega_L - \omega$.

The resulting transverse polarization close to resonance is, thus,
proportional to $\omega_1 / [ (\omega - \omega_L)^2 + \omega_1^2 (T_1/T_2) +
1/T_2^2 ]$ with the phenomenological relaxation rate $1/T_2$.
Thus, two relaxation terms are present, viz. the 'external' damping given by
$1/T_2^2$ and an intrinsic term $\omega_1^2 (T_1/T_2)$ given by the driving rf field itself.

Similarly, the same intrinsic mechanism should be expected if the driving
field $B_1$ is generated by a SOI-mediated bias like in the case considered in
the paper.  We thus anticipate a total spin relaxation rate of the
form\footnote{assuming $T_1 = T_2$ for simplicity} $\sqrt{\omega_R^2 +
  \Gamma^2}$ with Rabi frequency $\omega_R = e E_{0} \alpha / \hbar \omega_L$
derived at resonance from Eq.  (\ref{eq:spinbar}). Here, $E_0$ denotes the amplitude of the electric
field $\mathbf{E}(t) = E_0 \mathbf e_y \cos (\omega t)$ and $\Gamma$ is given
by Eq. (\ref{eq:damping2}). However, the Rabi frequency occurring in the rate
$\sqrt{\omega_R^2 + \Gamma^2}$, being E-field dependent, must be negligible
for a polarization $S^i$ which is calculated in linear response with respect
to $\mathbf{E}(t)$. This imposes the self-consistent condition
\begin{equation}
  \label{eq:lin_response}
  \omega_R \ll \Gamma \quad \Leftrightarrow  \quad \frac{\hbar e E_0}{p_F \omega_L
  \tau} \ll 2 \alpha p_F \left( 1 +  \frac{1}{2[1 + (\omega_L
 \tau)^2]}     \right) 
\end{equation}
for the validity of the linear response approach. A more systematic approach for estimating
the validity of the linear response regime requires an explicit evaluation
of the non-linear response, which, however, is beyond the scope of the present work.

Secondly, in order to carry out the momentum integrals in
Eq.(\ref{eq:spin-spin}) we introduced a condition limiting the SOI strength
\begin{equation}
  \label{eq:a}
  a =  \frac{\alpha p_F}{\hbar \omega_L} \ll 1 .
\end{equation}
This constraint not only simplified our analysis but also defines the most
interesting regime for EDSR. Indeed, in order to have a pronounced resonance,
the width of the resonance peak needs to be smaller than the resonance
frequency, i.e. $\Gamma\ll \omega_{L}$, which is equivalent to $\alpha p_F
x\ll \hbar \omega_L$ (see Eq. (\ref{eq:damping2})). For self-consistency we need to assume
$x \leq 1$ (see the text before Eq. (\ref{eq:mag3trans})), and thus we see that $a\ll 1$
ensures $\Gamma\ll \omega_{L}$.

In this context we note the somewhat
counterintuitive fact that the height of the resonance decreases with increasing
SOI, see Eq. (\ref{eq:mag3trans}).  Indeed, on one hand the polarization is proportional to 
$\alpha$ via the driving rf field, and thus increases with increasing SOI. 
On the other hand, at  resonance
the polarization becomes proportional to $1/\Gamma$ (due to disorder)
which gives then rise 
to a suppression factor $ 1/\alpha^{2}$. Thus, in total the polarization
decrease as $ 1/\alpha$ with increasing SOI at resonance.

Our last constraints
\begin{equation}
  \label{eq:fermi}
  \frac{b_0}{E_F}, \frac{\hbar \omega}{E_F},   \frac{\hbar}{E_F \tau} \ll 1
\end{equation}
correspond to the physically relevant situation where the Fermi energy $E_{F}$ is the
largest energy in the system. Further, the condition $x = 2 \alpha p_F
\tau/ \hbar \ll 1$ does not restrict the validity of Eqs. (\ref{eq:sigma1}) and (\ref{eq:damping}) 
but permits us to represent Eq. (\ref{eq:mag3trans}) in terms of  two Lorentzians. 
In the case $\omega \tau \approx 1$, however, it becomes equivalent
to the inequality (\ref{eq:a}).

\section{Numerical estimates}\label{sec:estimates}
\begin{table}[t]
  \centering
\setlength{\extrarowheight}{5pt}
  \begin{tabular}{|l|c|l|}
    \hline
    Description &   Parameter & Value \\ \hline \hline
    sheet density & $n_2$ & $4 \times 10^{11} \; \textrm{cm}^{-2}$  \\ \hline
    effective mass & $m^\ast$ & $0.067 \, m_e$ \\ \hline
    scattering time & $\tau$ & $2 \times 10^{-11} \, \textrm{s}$ \\ \hline
    frequency & $f = \omega / 2 \pi$ & $8 \; \mathrm{GHz}$ \\ \hline
    Larmor frequency & $f = f_L = \omega_L/ 2 \pi $ & $8 \; \mathrm{GHz}$  \\ \hline
    Rashba Parameter & $\alpha$ & $10^{-12} \;\mathrm{eV\, cm}$ \\ \hline
    electric field & E & $ 1.66 \; \mathrm{V \, cm}^{-1}$ \\ \hline 
    polarization & $P$ & $10^{-4}$ \\ \hline
    SOI vs. scattering & $x$ & $0.1$ \\ \hline
    spin relaxation rate & $\Gamma /2 \pi$ & $0.05 \; \mathrm{GHz}$ \\ \hline
    resonance shift & $\delta \omega /2 \pi$ & $0.01 \; \mathrm{GHz}$ \\ \hline 
    Rabi frequency & $\omega_R /2 \pi$ &  $0.012 \; \mathrm{GHz}$ \\ \hline \hline
    \multicolumn{3}{|c|}{validity conditions} \\ \hline 
    linear response & $\omega_R / \Gamma$ & $ 0.27 $ \\ \hline
    relative SOI strength & $a = \alpha p_F / \hbar \omega_L$ & $0.05$ \\ \hline
  \end{tabular}
  \caption{Numerical estimates}
  \label{tab:parameters}
\end{table}

To illustrate the predicted effects we now evaluate the polarization explicitly using typical GaAs parameters
 (cf. table \ref{tab:parameters}), thereby making sure that we stay within the
 range of validity of our approximations. With a typical sheet density $n_2 = p_F^2/ 2 \pi \hbar^2 = 4 \times
10^{15} \; \mathrm{m}^{-2}$, effective mass $m^\ast = 0.067 \, m_e$ and a
high mobility scattering time $\tau =
2 \times 10^{-11} \; \mathrm{s}$ taken
from \cite{beenakker} we can estimate the maximum polarization $P$ as the
ratio of the peak polarization per unit area and the sheet density
\begin{align}
  \label{eq:polarization}
P & = \frac{S^3_{max}}{n_2} = \frac{e E m^\ast}{(2 \pi \hbar n_2)^2 2
  \alpha \tau} \notag \\
& \times \frac{\omega \tau}{\sqrt{1 + \omega^2 \tau^2}} \left( 1 + \frac{1}{2 (1 +
    \omega^2 \tau^2)} \right)^{-1} \, \, .
\end{align}
In order to stay within the condition (\ref{eq:a}) we choose a small Rashba
- parameter $\alpha = 10^{-14} \; \mathrm{eV m}$ and find $x= 0.1$. Assuming
a realistic microwave frequency $\omega = 50 \; \times 10^{9} \; \mathrm{s}^{-1}$
corresponding to $\omega \tau =1$ and a voltage amplitude of $V = 0.1 \;
\mathrm{V}$ over a sample length of $l = 600 \; \mathrm{\mu m}$ we find an
electric field $E_0 = 166 \; \mathrm{V m^{-1}}$ and a polarization of
\begin{equation}
  \label{eq:polarization1}
  P = 10^{-4}\,\, .
\end{equation}
Note that the size of the chosen $E-$field satisfies the linear response condition
(\ref{eq:lin_response}) (and poses no severe limitation for a real experiment).

The corresponding number of excess spins $N_\uparrow - N_\downarrow$ in a
laser spot of size $5\mu m \times 5 \mu m$ is $200$. This number is measurable
with state-of-the-art optical  detection techniques such as Faraday rotation\cite{awschalom-she-2deg}.

We can further quantify the peak width $\Gamma$ and the frequency shift $\delta \omega$. 
Making use of Eqs. (\ref{eq:damping2}) and (\ref{eq:damping1}) of the main text we
find
\begin{align}
  \label{eq:numshift}
\Gamma &= 0.3 \times 10^{9} \; \mathrm{s}^{-1} \notag \\
\delta \omega &=  0.06 \times 10^{9} \; \mathrm{s}^{-1}  .
\end{align}
As a further characterization of the resonance we estimate the Rabi frequency
$\omega_{R}$, given by the amplitude of $b_1(t)$ in Eq. (\ref{eq:spinbar}). Assuming a bias
$\mathbf{E}(t) = E_0 \mathbf e_y \cos (\omega t)$ we find
\begin{equation}
  \label{eq:rabi}
  \omega_R = \frac{e E_0 \alpha}{\hbar \omega} = 0.08 \times 10^{9} \; \mathrm{s}^{-1},
\end{equation}
evaluated at resonance\footnote{corresponding to a magnetic field $B \approx 1
  \; \mathrm{T}$ for $|g| =0.44$} $\omega_L = \omega$ with the parameters
given above. A summary of the above calculation and a check of the constraints
Eqs. (\ref{eq:a},\ref{eq:lin_response}) is given in {table
  \ref{tab:parameters}}.

\section{Spin Hall current and polarization}\label{sec:spin-current}
We show now that the obtained polarization (magnetization/$\mu_{B}$) $S^i$ can be related to the spin
current (defined below) via an exact relation.  More generally, we consider the
spin density operator
\begin{equation}
  \label{eq:spin-density}
  \rho^{i}(\mathbf x) =
\frac{1}{2}  \{\sigma^{i} , \delta(\mathbf x -
\hat{ \mathbf x})\},
\end{equation}
 defined as the (symmetrized) product of the spin with the
particle density operator $\delta (\mathbf x - \mathbf{\hat{x}})$ where
$\hat{\mathbf{x}}$ is the  position operator. 
Integrating over space (homogeneous limit) and taking expectation values we get the spin polarization $S^i = \int d^2x <\rho^i(\mathbf  x)>$.
The spin current density associated
with $\rho^i$ is defined in the usual way\cite{sinova,schliemann-spin, inoue,mishenko,dimitrova, erlingsson,oleg}
\begin{equation}
\label{eq:spincurrent0}
j^{i}_k(\mathbf x,t) =  \frac{1}{2} \{ \sigma^{i}, j_k(\mathbf x) \} 
\end{equation}
in terms of the current operator $j_k(\mathbf x,t) = \frac{1}{2} \{ \delta(x-
\hat{ \mathbf x} ) , v_k \}$ where, in contrast to the linear response treatment of
the paper, the velocity operator $v_k = i/\hbar [H, x_k] =  (p_k - (e/c) A_k)/m + \alpha  
 (  \boldsymbol{\sigma} \times  \mathbf e_z)_k $ contains the kinetic momentum
  including the (homogenous) vector potential $\mathbf A$.

The two operators
  $\rho^i$ and $j^i_k$ are related via the
 Heisenberg equation of motion
\begin{equation}
\label{eq:eq_mot}
\frac{d}{dt} \rho^{\eta}(\mathbf x,t) = \frac{i}{\hbar} [H, \rho^{\eta}]
\end{equation}
given by the Hamiltonian Eq. (\ref{eq:h_ext}). Analogous to \cite{erlingsson} where the
Rashba- and Dresselhaus SOI has been considered it forms an
exact operator identity 
\begin{align}
\label{eq:continuity}
\frac{d}{dt} \rho^1(\mathbf x,t) + \nabla \cdot \mathbf j^1(\mathbf x,t)&= -
\frac{2 \alpha m}{\hbar} j^3_x(\mathbf x,t) \notag  \\ &- \frac{2}{\hbar} \left[
  \rho^2(\mathbf x,t) b_{0,z} - \rho^3(\mathbf
  x,t) b_{0,y} \right]  \notag \\
\frac{d}{dt} \rho^2(\mathbf x,t) + \nabla \cdot \mathbf j^2(\mathbf x,t)&= -
\frac{2 \alpha m}{\hbar} j^3_y(\mathbf x,t) \notag \\ &- \frac{2}{\hbar} \left[
  \rho^3(\mathbf x,t) b_{0,x} - \rho^1(\mathbf
  x,t) b_{0,z} \right] \nonumber \notag \\
\frac{d}{dt} \rho^3(\mathbf x,t) + \nabla \cdot \mathbf j^3(\mathbf x,t)&= +
\frac{2 \alpha m}{\hbar}  \left[ j^1_x(\mathbf x,t) + j^2_y(\mathbf x,t)
\right] \notag \\ &-
\frac{2}{\hbar} \left[\rho^1(\mathbf x,t) b_{0,y} - \rho^2(\mathbf x,t)
  b_{0,x} \right] ,
\end{align}
for the case of an additional static magnetic field with components $b_{0,i}, \;i=x,y,z$,
which holds independently of the impurity potential as $\rho^i$ commutes with
the position operator.

In deriving Eq. (\ref{eq:continuity}) the definition of
$j^i_k$ arises naturally as a divergence term of a current
associated with the spin density. Together with the time derivative $\dot
\rho^i$ it forms the left-hand side of a continuity equation. The right hand
side, however, is nonzero and describes the dynamics of the spin due to
the external magnetic field $\mathbf b_0$ and the internal SOI field. The definition of
Eq.(\ref{eq:spincurrent0}) as a 'spin current' is thus ambiguous\cite{rashba-equilibrium-sc,engel}
and it is not clear to what extent the quantity Eq. (\ref{eq:spincurrent0}) can
be identified with actual spin transport, i.e. with  spin polarized currents which are
experimentally accessible\cite{erlingsson2}.

In spite of the above concerns we note that in the homogeneous limit the 
spin Hall current
can be expressed entirely in terms  of the polarization. Namely, going over
to the spin Hall current $I^i_k = \int d^2 x< j^i_k (x)>$ such that the gradient
in Eq.(\ref{eq:continuity}) vanishes we find the expectation value of the  spin Hall current
given by
\begin{align}
  \label{eq:spincurrent}
  I^3_x(\omega) &= \frac{\hbar}{2 \alpha m } \left[ i \omega S^1 (\omega) +
  \omega_{L}S^3(\omega) \right] = \frac{e}{2 \pi \hbar} E_2(\omega)
  \notag \\
& \times \bigg[ i \omega \tau \left(1 -
  \Sigma^{11} \left(1- \frac{1}{\lambda} \right) \right)
- \omega_L \tau \Sigma^{31} \left(1- \frac{1}{\lambda} \right) \bigg].
\end{align} 
[This relation can be obtained directly from the Heisenberg equation of motion
$d {\sigma}^{1}/dt=i[H,{\sigma}^{1}]$, and by
noting that $ I^3_x=\sigma^{3}p_{x}$.]
Since $S^3$ vanishes for $\omega = 0$ it is obvious from Eq.
(\ref{eq:spincurrent}) that there is no spin Hall current in the dc  limit
$\omega \to 0$ for a homogenous infinite sample\cite{inoue,erlingsson,oleg}.
This means a generalization of the argument given in \cite{erlingsson,oleg} to
the case of a finite magnetic field.  For finite frequencies, however, Eq.
(\ref{eq:spincurrent}) predicts a non-vanishing oscillating spin current
expressed in terms of the polarization components perpendicular to the
applied electric rf field. With the results for $S^i$ inserted we find the ac
spin Hall conductivity evaluated at resonance ($\omega=\omega_{L}$) as
\begin{equation}
  \label{eq:sh-conductivity}
  \sigma^{3,res}_{x y}  \equiv \frac{ I^3_x\hbar/2}{E_y} =\frac{e}{4 \pi} \frac{i \omega_L \tau}{1 + 2 \lambda(\omega_L)} .
\end{equation}
We emphasize that this relation provides a direct link between the experimentally accessible
polarization and the spin current. For $\omega_L \tau \gg 1$ this becomes
\begin{equation}
  \label{eq:sh-conductivityRes}
  \sigma^{3, res}_{x y}= - \frac{e}{8 \pi}
\end{equation}
giving a universal value for the spin current at resonance. It is quite
remarkable that the same result, Eq. (\ref{eq:sh-conductivityRes}), can be
obtained when inserting the solutions $S^{i}$ obtained from the Bloch
equations Eq. (\ref{eq:bloch}) close at resonance into Eq.
(\ref{eq:spincurrent}).


\end{document}